\documentclass[12pt, letterpaper]{article}
\usepackage{amsmath, amssymb,color, natbib, graphicx, caption, subfig, titlesec, setspace, authblk, epstopdf}
\usepackage[letterpaper, margin=1in]{geometry}

\titleformat*{\section}{\large\bfseries}
\title{Modelling extremes using\\ approximate Bayesian Computation}
\date{}
\author{R. Erhardt\footnote{Wake Forest University, U. S. A.}\: and S. A. Sisson\footnote{School of Mathematics and Statistics, University of New South Wales}}

\begin{document}
\maketitle

\begin{abstract} 
\noindent By the nature of their construction, many statistical models for extremes result in likelihood functions that are computationally prohibitive to evaluate. This is consequently problematic for the purposes of likelihood-based inference. With a focus on the Bayesian framework, this chapter examines the use of approximate Bayesian computation (ABC) techniques for the fitting and analysis of statistical models for extremes. After introducing the ideas behind ABC algorithms and methods, we demonstrate their application to extremal models in stereology and  spatial extremes.
\end{abstract}

%%%%%%%%%%%
%%%%%%%%%%%%
 \section{Introduction}%%
 %%%%%%%%%%%%
 %%%%%%%%%%%
\label{sec:intro}

Suppose interest is in modelling the extremes of a multivariate random process.  A useful example to hold in mind might be measurements of temperature $y$ sampled at locations $x_1, ..., x_D$, where there is dependence among the $D$ locations due to their proximity to one another.  The extremal dependence may, in general, differ from the dependence of non-extremes, and so the model should target the extremes only and not allow the bulk of non-extreme data to overwhelm to model fit.  Models for extremes are useful when trying to estimate the risks associated with rare but influential events.

Call $y_{obs}$ the observed data from a random process $f(y | \theta)$, where $\theta$ is the parameter, and call $\mathcal{A}$ an event or set of events of particular concern.  The probability of such an event (with notation to remind of the influence from the parameter $\theta$) is
\[
p_{\theta} = Pr(y_{obs} \in A) = \int_{\mathcal{A}} f(y | \theta) dy,
\]
and  $p^{-1}_{\theta}$ is termed the \textit{return period} of the event $\mathcal{A}$.  A return period of twenty years means that the event occurs roughly once every twenty years.  Estimating return periods of extremes is crucial to designing systems capable of handling such extremes.  Examples include flood walls that can withstand the 100-year flood, infrastructure that can withstand extreme heat or cold, insurance companies that can remain solvent after losses of a particular magnitude, and so forth.  Given the sparsity of data on extreme events, information on $\theta$ can be minimal, and this can translate into large parameter uncertainty in $\theta$.  \citet{coles1996bayesian} and {\bf Stephenson (earlier chapter in this book)} advocate the Bayesian approach to inference, arguing that expert opinion incorporated through a prior distribution $p(\theta)$ could be of tremendous value given the natural scarcity of observed data extremes.  Alternatively, if one chooses a vague prior $p(\theta)$ and obtains a posterior $\pi(\theta | y_{obs})$, then the parameter uncertainty in $\theta$ would naturally be incorporated into calculations, such as predictive return levels.

Now, consider models for multivariate extremes.  Suppose that $Y_{1}, ..., Y_{n}$ are univariate, independent and identically distributed replicates from some distribution function $F$, and define $M_{n} = \mbox{max}(Y_{1}, ..., Y_{n})$ as the maximum of the $n$ random variables.  The distribution of $M_{n}$ can be obtained exactly assuming $F$ is known.  In practice, then one could estimate $F$ from all of the data $Y_{1}, ..., Y_{n}$ and estimate the distribution of $M_{n}$ as $P(M_{n} \leq z) = F^{n}(z)$, but this approach has two drawbacks.  The first is that even minor discrepancies in estimating $F$ result in large discrepancies in $F^{n}$, particularly in the tails of $F$.  Put another way, why should a model which fits the bulk of data to also be a good fit in the tails?  A second drawback is that in the limit as $n \rightarrow \infty$, $F^{n}$ does not converge to a non-degenerate distribution.  Instead, one may model renormalized maxima $\left(M_{n} - b_{n} \right)/a_{n}$ for sequences $a_{n} > 0$ and $b_{n}$.  If there exist sequences $a_{n} > 0$ and $b_{n}$ such that
\begin{equation*}
\lim_{n \rightarrow \infty} P\left(\frac{M_{n} - b_{n}}{a_{n}} \leq z \right) \rightarrow G(z)
\end{equation*}
for some non-degenerate distribution function $G$, then $G$ is a member of the \textit{Generalized Extreme Value} (GEV) family, with distribution function
\begin{equation}
G(z) = \exp \left[-\left(1+ \xi \frac{z - \mu}{\sigma}\right)_{+}^{-1/\xi} \right].
\label{eq:gev}
\end{equation}
Here $a_{+} = \mbox{max}(a,0),$ and $\mu, \sigma, $ and $\xi$ are the location, scale, and shape parameters, respectively \citep{coles2001introduction, gnedenko1944limit}.  The sign of the shape parameter $\xi$ determines that $G$ corresponds to one of the three classical extreme values distributions: $\xi>0$ is Fr\'{e}chet with support $z \in [\mu - \sigma/\xi, + \infty)$, $\xi<0$ is Weibull with support $z \in (-\infty, \mu - \sigma/\xi]$, and $\xi \rightarrow 0$ is Gumbel with support $z \in (-\infty, +\infty)$.  The Generalized Extreme Value distribution $G$ has the property of max-stability, understood as follows: if $Y_{1}, ..., Y_{n}$ are i.i.d. draws from $G$, then $\mbox{max}(Y_{1}, ..., Y_{n})$ also has distribution $G$, meaning
\begin{equation*}
G^{n}(A_{n}z + B_{n}) = G(z)
\end{equation*}
for appropriate sequences $A_{n} > 0$ and $B_{n}$.  In fact, a distribution is max-stable if and only if it is a member of the GEV family \citep{leadbetterrootz}.  If block maxima are taken over a block size large enough to allow the GEV to be a valid approximation, then if one further increased the block size (from monthly to annual maxima, for example) the GEV model would still hold, with only a change in the three parameters.  While these results for the GEV family assume i.i.d. data, this assumption can be relaxed and the limiting distribution still holds so long as certain mixing conditions are satisfied \citep{leadbetterrootz}.

A useful member of the GEV family is the unit-Fr\'{e}chet distribution, with distribution function
\begin{equation*}
P(Z \leq z) = \exp \left( -{1\over z} \right).
\end{equation*}
The simplicity of the unit-Fr\'{e}chet  distribution function is helpful when one considers multivariate and ultimately spatial extremes.  Any member of the GEV family may be transformed to have unit-Fr\'{e}chet margins as follows: if $Z$ has a GEV distribution, and a new variable $U$ is defined as
\begin{equation}
U = \left(1 + \xi\frac{Z - \mu}{\sigma}\right)^{1/\xi},
\label{eq:u}
\end{equation}
then $U$ has unit-Fr\'{e}chet margins.  This transformation assumes that the parameters are known.  If the parameters are unknown, they may first be estimated and then the transformation to $U$ is taken.  For extreme values data in a spatial setting, the first step is often to transform data at each location to unit-Fr\'{e}chet by fitting all marginal distributions.  Then the second step is to analyze the spatial dependence among sites once every location has been transformed.  Thus, there is no loss of generality when one assumes unit-Fr\'{e}chet margins.

This approach can be extended to handle multivariate extremes.  Let $(X_{i 1}, ..., X_{i D})$, $i=1, ..., n$ be a $D-$dimensional random vector and let $M_{n} = (M_{n 1}, ..., M_{n D})$ be the vector of componentwise maxima, where $M_{n d} = \max(X_{1 d}, ..., X_{n d})$ for $d=1, ..., D$.  It is worth noting that $M_{n}$ will not appear in the data record unless the occurrence times of each element's block maximum happen to coincide.  In a spatial context, this vector $M_{n}$ might refer to the annual maxima of some variable at $D$ locations.  A non-degenerate limit for $M_{n}$ exists if there exist sequences $a_{nd} > 0$ and $b_{nd}$, $d=1, ..., D$ such that
\begin{equation*}
\lim_{n \rightarrow \infty} P \left(\frac{M_{n1}-b_{n1}}{a_{n1}} \leq z_{1}, \ldots , \frac{M_{nD}-b_{nD}}{a_{nD}} \leq z_{D} \right) = G(z_{1}, \ldots , z_{D}).
\end{equation*}
Then $G$ is a multivariate extreme value distribution (MEVD), and is max-stable if there exist sequences $A_{nd}>0$, $B_{nd}$, $d=1, ..., D$ such that, for any $n>1$
\begin{equation*}
G^{n}(z_{1}, \ldots , z_{D}) = G(A_{n1}z_{1} + B_{n1}, \ldots , A_{nD}z_{D} + B_{nD}).
\end{equation*}

The univariate marginal distributions of a multivariate extreme value distribution are all necessarily GEV distributions.  Thus, for each margin one can define a transformation with parameter $(\mu_d, \sigma_d, \xi_d)$ and transform to unit-Fr\'{e}chet using equation (\ref{eq:u}).  Since all GEV distributions can be transformed into unit-Fr\'{e}chet, all MEVD can be transformed into multivariate unit-Fr\'{e}chet, and thus one may assume, without loss of generality, that all MEVD have unit Fr\'{e}chet margins.  This works out because the domain of attraction condition is preserved under monotone transformations of the marginal distributions \citep{resnickextreme}.  For $D$ fixed locations $x_1,\ldots,x_D$, the joint distribution function can be written as
\begin{equation}
P(Z(x_{1}) \leq z_{1}, \ldots , Z(x_{D}) \leq z_{D}) = \exp\left(-V(z_{1}, \ldots , z_{D})\right)
\label{eq:v}
\end{equation}
where $V(z_{1}, ..., z_{D})$ is the exponent measure first described by \citet{pickands1981multivariate}.  This function takes the form
\begin{equation}
V(z) = D \cdot \int_{\Delta_D} \max_{d=1, \ldots , D}\frac{w_d}{z_d}H(\,dw)
\label{eq:Vsolve}
\end{equation}
where $\Delta_D = {w \in \mathbb{R}^{D}_{+} \mid w_1 + \ldots + w_D =1}$ is the $D-1$ dimensional simplex, and the angular (or spectral) measure $H$ is a probability measure on $\Delta_D$ which determines the dependence structure of the random vector.  Due to the common marginal distributions, $H$ has moment conditions $\int_{\Delta_D} w_dH(\,w) = 1/D$ for $d=1, ..., D$.  Max-stability implies that for all $N$,
\begin{eqnarray*}
P(Z_{1} \leq z_{1}, \ldots , Z_{D} \leq z_{D})^{N} & = & \exp(-N\cdot V(z_{1}, \ldots , z_{D}))\\
& = & \exp(-V(z_{1}/N, \ldots , z_{D}/N))
\end{eqnarray*}
with the final equality following from the homogeneity property of the exponent measure.  The measure also satisfies two bounds: if all locations are independent, $V(z_{1}, ..., z_{D}) = 1/z_{1} + ... + 1/z_{D}$; if all locations are totally dependent, $V(z_{1}, \ldots , z_{D}) = \mbox{max}(1/z_{1}, \ldots , 1/z_{D})$.  Thus, we always have
\[
\mbox{max}(1/z_{1}, \ldots , 1/z_{D}) \leq V(z_{1}, \ldots,  z_{D}) \leq 1/z_{1} + \ldots + 1/z_{D}.
\]

There are two challenges to working with the spectral representation of the joint distribution function shown in equation (\ref{eq:v}).  First, even if a closed form for the exponent measure can be found by solving equation (\ref{eq:Vsolve}), the joint density function undergoes a combinatorial explosion as the dimension $D$ increases.  Differentiating $\exp(-V)$ with respect to the values $z_{1}, ..., z_{D}$ leads to a rapid growth in terms:
\begin{eqnarray*}
\frac{\delta}{\delta z_1}\exp(-V) & = & -V_{1} \exp(-V)\\
\frac{\delta^2}{\delta z_1z_2}\exp(-V) & = & (V_{1}V_{2} - V_{12}) \exp(-V)\\
\frac{\delta^3}{\delta z_1z_2z_3}\exp(-V) & = & (-V_{1}V_{2}V_{3} + V_{12}V_{3} + V_{13}V_{2} + V_{23}V_{1} - V_{123}) \exp(-V)\\
&& \ldots
\end{eqnarray*}
where $V_{i}$ is the partial derivative of V with respect to $z_{i}$.  Thus even if a reasonable choice for $V$ can be found, as the dimension $D$ increases one is left with an unwieldy likelihood function, which may be difficult to maximize.  More common, though, is the situation where closed-form expressions for the exponent measure cannot be obtained by solving equation (\ref{eq:Vsolve}).

As a result, the lack of a closed-form likelihood presents a stumbling block for modelling high dimensional extremes data. While there are a number of procedures that may permit some form of statistical inference in this setting (see the other chapters in this Handbook), the remainder of this chapter will demonstrate how approximate Bayesian computing can be implemented as one solution to this problem in the Bayesian framework.

%%%%%%%%%%%%%%%%%%%%%%%%%%%%
%%%%%%%%%%%%%%%%%%%%%%%%%%%%%
\section{A primer on approximate Bayesian computation}%%
%%%%%%%%%%%%%%%%%%%%%%%%%%%%%
%%%%%%%%%%%%%%%%%%%%%%%%%%%%
\label{sec:ABCbasics}

Suppose that interest is in performing a standard Bayesian analysis for a parameter $\theta\in\Theta$ which has a prior distribution $p(\theta)$. The model is defined through the likelihood function $L(y|\theta)$, which is assumed to be a candidate for the true data generating process that produced the observed dataset $y_{obs}\in{\mathcal Y}$. The posterior distribution of the model parameter $\theta$, having now observed the data $y_{obs}$ through the likelihood function, is expressed as $\pi(\theta|y_{obs})\propto L(y_{obs}|\theta)p(\theta)$. In the present setting, we could have e.g. $\theta=(\mu,\sigma,\xi)$ as the parameters of a generalised extreme value distribution.  The posterior distribution contains all the information that is needed to perform inference on the model (see e.g. \citet{gelman+csr03,ohagan+f04}).

As the posterior distribution is rarely available in closed form, it is common to base subsequent analysis on a Monte Carlo approximation to the posterior \citep{brooks+gjm11}. In this manner, if $\theta^{(1)},\ldots,\theta^{(N)}\sim\pi(\theta|y_{obs})$ are $N$ samples drawn from the posterior distribution, then the posterior expectation of some function $a(\theta)$ can be approximated by
\[
	{\mathbb E}_\pi[ a(\theta) ] = \int_\Theta a(\theta)\pi(\theta|y_{obs})d\theta \approx \frac{1}{N}\sum_{i=1}^N a(\theta^{(i)}),
\]
where ${\mathbb E}_\pi$ denotes expectation under $\pi$. There are a wide variety of algorithms available to draw samples from the posterior distribution \citep{doucet+fg01,chen+si00,brooks+gjm11}. One of the simplest of these is importance sampling, as illustrated in Algorithm \ref{alg:importanceSampling}, which produces weighted samples $(w^{(1)},\theta^{(1)}),\ldots,(w^{(N)},\theta^{(N)})$ from $\pi(\theta|y_{obs})$ based on samples $\theta^{(1)},\ldots,\theta^{(N)}$ from a sampling distribution $g(\theta)$. In this setting, and writing $w(\theta)=\pi(\theta|y_{obs})/g(\theta)$, then
\begin{eqnarray*}
	{\mathbb E}_g[ w(\theta)a(\theta) ] & = &\int_\Theta w(\theta) a(\theta)g(\theta)d\theta
	=\int_\Theta a(\theta)\pi(\theta|y_{obs})d\theta={\mathbb E}_\pi[a(\theta)]\\
	& \approx &\sum_{i=1}^N w(\theta^{(i)})a(\theta^{(i)}).
\end{eqnarray*}
That is, appropriately weighted samples from $g(\theta)$  can be used as samples from $\pi(\theta|y_{obs})$.

\begin{table}[htb]
\begin{tabular}{l}
%\vspace{1mm}
%\rule{\textwidth}{0.1mm}
\hline
\underline{Simple importance sampling algorithm}
\vspace{3mm}
\\

{\bf Input:} \\An observed dataset, $y_{obs}$. \\A desired number of samples $N>0$. \\A sampling distribution $g(\theta)$, with $g(\theta)>0$ if $p(\theta)>0$.
\vspace{3mm}
\\

{\bf Iterate:} For $i=1,\ldots,N$:\\
%\begin{enumerate}
%\item
1. Sample a parameter vector from sampling distribution $\theta^{(i)}\sim g(\theta)$.\\
%\item 
2. Weight each sample $\theta^{(i)}$ by $w^{(i)}\propto \pi(\theta^{(i)}|y_{obs})/g(\theta^{(i)})$.
\vspace{3mm}\\
%\end{enumerate}

{\bf Output:}\\ A set of $N$ weighted samples $(w^{(1)},\theta^{(1)}),\ldots,(w^{(N)},\theta^{(N)})$ drawn from $\pi(\theta|y_{obs})$.\\
\hline
%\rule{\textwidth}{0.1mm}
\end{tabular}
\caption{\small A simple importance sampling algorithm, based on a single large sample of size $N$. \label{alg:importanceSampling}}
\end{table}

Almost all posterior simulation algorithms need to be able to evaluate the likelihood function $L(y|\theta)$ in order to be correctly implemented. In the importance sampling algorithm (Algorithm \ref{alg:importanceSampling}) this occurs in the weight evaluation $w^{(i)} \propto L(y_{obs}|\theta^{(i)})p(\theta^{(i)})/g(\theta^{(i)})$. In the present setting, the natural construction of many useful statistical models for extremes results in the likelihood function being computationally prohibitive to evaluate (see Section \ref{sec:intro}). This computational intractability means that for these classes of models, an alternative procedure is needed to sample from the posterior distribution {\em without} directly evaluating the likelihood function.  One class of procedures that has been developed to achieve this is known as approximate Bayesian computation \citep{beaumont10,bertorelle+bm10,csillery+bgf10,sisson+f11,marin+prr12}.

%%%%%%%%%%%%%%%%%%%%%%%%%%%%
%%%%%%%%%%%%%%%%%%%%%%%%%%%%%
\subsection{Approximate Bayesian computation basics}%%
%%%%%%%%%%%%%%%%%%%%%%%%%%%%%
%%%%%%%%%%%%%%%%%%%%%%%%%%%%

All approximate Bayesian computation (ABC) procedures operate on the following heuristic argument, that was first developed in the population genetics literature \citep{tavare+bgd97,pritchard+spf99}. Suppose that we have a parameter vector $\theta^{(i)}$ that is a candidate draw from the posterior distribution $\pi(\theta|y_{obs})$. Further suppose that we can quickly generate an auxiliary dataset from the model, conditional on this candidate parameter vector, so that $y^{(i)}\sim L(y|\theta^{(i)})$. Now, the argument states that if $y^{(i)}$ and $y_{obs}$ are ``close'' to each other (in a general sense that will be made more precise below), then it is credible that the parameter vector $\theta^{(i)}$ could also have generated the observed dataset $y_{obs}$. In which case, the parameter vector $\theta^{(i)}$ should be retained as an approximate sample from $\pi(\theta|y_{obs})$. Conversely, if $y^{(i)}$ and $y_{obs}$ are not ``close''  to each other, then $\theta^{(i)}$ is unlikely to be able to generate the observed dataset, and so it should be discarded as not being a draw from the posterior. Repeating this procedure will produce samples that are either approximate draws from the posterior $\pi(\theta|y_{obs})$, or exact draws from some as yet unspecified approximation to the posterior distribution. Either way, direct numerical evaluation of the computationally intractable likelihood function has been avoided.

The principles behind the above heuristic method can be made more precise. Suppose that we generate our candidate parameter vectors from the prior $\theta^{(i)}\sim p(\theta)$. (As part of an importance sampling algorithm, the candidate parameter vectors $\theta^{(i)}$ may be generated from $g(\theta)$ and then reweighted.) Then, given this, a dataset is generated from the likelihood. In this way, the pair
\[
	(\theta^{(i)},y^{(i)}) \sim L(y|\theta)p(\theta)
\]
has been generated from the prior predictive distribution $p(y,\theta)=L(y|\theta)p(\theta)$.
In general, the auxiliary and observed dataset can be compared via some distance metric $\|y^{(i)}-y_{obs}\|$, such as Euclidean or Mahalanobis distance. The retention and discarding of ``close'' and not ``close'' auxiliary datasets can be mimicked by computing an importance weight $K_h(\|y^{(i)}-y_{obs}\|)$ where $K_h(u)=K(u/h)/h$ is a standard smoothing kernel with scale parameter $h>0$. For example, if $K_h$ corresponds to the uniform kernel over $(-h,h)$, then $\theta^{(i)}$ will receive the weight $1$ if $\|y^{(i)}-y_{obs}\|\leq h$ (i.e. it is retained if $y^{(i)}$ and $y_{obs}$ are sufficiently ``close'') and the weight 0 if $\|y^{(i)}-y_{obs}\|> h$ (i.e. it is rejected). Other choices of kernel $K_h$ will produce continuous weights.

The resulting weighted samples $(w^{(i)},\theta^{(i)},y^{(i)})$ are then draws from
\begin{equation}
	\label{eqn:ABCjointPost}
	\pi_h^{ABC}(\theta,y|y_{obs}) \propto K_h(\|y-y_{obs}\|)L(y|\theta)p(\theta)
\end{equation}
\citep{reeves+p05,wilkinson08}.
Distribution (\ref{eqn:ABCjointPost}) is the joint posterior distribution of model parameter $\theta$ and auxiliary dataset $y$ such that $y$ and $y_{obs}$ are ``close'' in a specific sense. If we are just interested in the resulting distribution of the parameter vector $\theta$ as an approximation of the posterior distribution $\pi(\theta|y_{obs})$, then
\begin{equation}
	\label{eqn:ABCpost}
	\pi_h^{ABC}(\theta|y_{obs}) \propto  \int_{\mathcal Y} K_h(\|y-y_{obs}\|)L(y|\theta)p(\theta) dy.
\end{equation}
Equation (\ref{eqn:ABCpost}) is the ABC approximation to the true posterior distribution $\pi(\theta|y_{obs})$.

The quality of this approximation is determined by the kernel scale parameter $h>0$. Consider what happens to $\pi_h^{ABC}(\theta|y_{obs})$ as $h$ gets small. We have
\begin{eqnarray*}
	\lim_{h\rightarrow0} \pi_h^{ABC}(\theta|y_{obs})
	&\propto & \lim_{h\rightarrow0} \int_{\mathcal Y} K_h(\|y-y_{obs}\|)L(y|\theta)p(\theta) dy\\
	& = & \int_{\mathcal Y} \boldsymbol{1}(y=y_{obs}) L(y|\theta)p(\theta) dy\\
	& = & L(y_{obs}|\theta)p(\theta)\\
	& \propto & \pi(\theta|y_{obs}),
\end{eqnarray*}
where $\boldsymbol{1}(A)=1$ if $A$ is true, and $0$ otherwise. That is, if we only accept those candidate parameter values $\theta^{(i)}$ that exactly reproduce the observed dataset, then we will exactly recover the true posterior distribution $\pi(\theta|y_{obs})$. In practice however, unless $y_{obs}$ is discrete, this can never occur (and in fact, will be unlikely to occur practically for discrete $y_{obs}$ in all but trivial analyses). Hence $h$ will typically be greater than zero in practice.

In the other extreme
\begin{eqnarray*}
	\lim_{h\rightarrow\infty} \pi_h^{ABC}(\theta|y_{obs})
	&\propto & \lim_{h\rightarrow\infty} \int_{\mathcal Y} K_h(\|y-y_{obs}\|)L(y|\theta)p(\theta) dy\\
	& \propto & \int_{\mathcal Y}  L(y|\theta)p(\theta) dy\\
	& = & p(\theta),
\end{eqnarray*}
assuming that $K_h(u)\propto 1$ as $h\rightarrow\infty$.
That is, sampling from the prior $p(\theta)$ and then not showing any discrimination in favour of auxiliary data closely matching the observed data will simply result in all samples being equally weighted, and the ABC approximation to the posterior distribution being given by the prior $p(\theta)$. It should be clear that increasing fidelity toward reproducing the observed data, as measured by decreasing $h$, defines a smooth transition from prior $p(\theta)$ to posterior $\pi(\theta|y_{obs})$, and that in practice the actually attainable ABC posterior approximation $\pi_h^{ABC}(\theta|y_{obs})$ lies somewhere between the two. Given that it results in greater closeness to $\pi(\theta|y_{obs})$, for inferential purposes lower $h$ is desirable.

Aside from the approximation to the posterior linked to $h$, in practice a second level of approximation is commonly introduced in an ABC analysis, and one which can have a greater impact on the quality of the ABC approximation of $\pi(\theta|y_{obs})$ than $h$. To understand the motivation for this, consider the form of the likelihood component of $\pi_h^{ABC}(\theta|y_{obs})$ (\ref{eqn:ABCpost}) where, for simplicity, both $y_{obs}$ and $\theta$ are univariate. Using a Taylor expansion and the substitution $u=(y-y_{obs})/h$, then
\begin{eqnarray*}
\int_{\mathcal Y} K_h(\|y-y_{obs}\|)L(y|\theta) dy & \approx &
L(y_{obs}|\theta) + \frac{1}{2}h^2L''(y_{obs}|\theta)\int u^2 K(u) du,
\end{eqnarray*}
assuming the usual kernel function properties of $\int K(u)du=1$, $\int uK(u)du=0$ and $K(u)=K(-u)$. That is, the ABC simulation procedure is simply performing a form of conditional kernel density estimation, by estimating a smoothed likelihood function, and then using this as part of a regular Bayesian analysis \citep{blum10}.

Under this interpretation, a limitation of the proposed ABC method becomes apparent. Kernel density estimation is well known to suffer from the curse of dimensionality, and is arguably impractical in more than two dimensions. Here, the appropriate dimensionality is in $y-y_{obs}$ (although it is masked by the univariate measure $\|y-y_{obs}\|$) within the kernel function $K_h$. That is, as the dimension of $y_{obs}$ increases, the performance of the ABC posterior approximation $\pi_h^{ABC}(\theta|y_{obs})$ will rapidly deteriorate \citep{blum10}. In effect, it becomes increasingly unlikely to be able to reproduce the observed dataset $y_{obs}$ by randomly sampling $y$, as the dimension of $y_{obs}$ increases, thereby forcing the practitioner to increase $h$ for a fixed number of samples $N$.

A simple solution to this is to reduce the dimensionality of $y_{obs}$ by reducing it to a vector of summary statistics $s_{obs}=S(y_{obs})$ and then performing the ABC procedure as before, but using $s_{obs}$ rather than $y_{obs}$ \citep{tavare+bgd97,pritchard+spf99,beaumont+zb02}. In this manner, (\ref{eqn:ABCjointPost}) and (\ref{eqn:ABCpost}) become
\begin{eqnarray}
\label{eqn:ABCjointPostSS}
\pi_h^{ABC}(\theta,s|s_{obs}) &\propto & K_h(\|s-s_{obs}\|)L(s|\theta)p(\theta)\\
\label{eqn:ABCpostSS}
\pi_h^{ABC}(\theta|s_{obs}) &\propto & \int_{\mathcal S}K_h(\|s-s_{obs}\|)L(s|\theta)p(\theta)ds,
\end{eqnarray}
where ${\mathcal S}=\{S(y): y\in{\mathcal Y}\}$ is the image of ${\mathcal Y}$ under $S$, and where $L(s|\theta)$ corresponds to the likelihood function of the summary statistic, which is also assumed to be computationally intractable. In the case where $S(y)=y$ then (\ref{eqn:ABCjointPostSS}) and (\ref{eqn:ABCpostSS}) reduce to (\ref{eqn:ABCjointPost}) and (\ref{eqn:ABCpost}).

Note that once a decision has been made on the summary statistics, then the most accurate possible ABC posterior approximation is given by $\pi(\theta|s_{obs})$ as
\begin{eqnarray*}
	\lim_{h\rightarrow0} \pi_h^{ABC}(\theta|s_{obs})
	&\propto & \lim_{h\rightarrow0} \int_{\mathcal S} K_h(\|s-s_{obs}\|)L(s|\theta)p(\theta) ds\\
	& = & \int_{\mathcal S} \boldsymbol{1}(s=s_{obs}) L(s|\theta)p(\theta) ds\\
	& = & L(s_{obs}|\theta)p(\theta)\\
	& \propto & \pi(\theta|s_{obs}),
\end{eqnarray*}
in the same manner as before. Consequently it is important that due consideration is given to the summary statistics aspect of the model.

In the case of sufficient statistics, there is no loss of information and so $\pi(\theta|s_{obs})=\pi(\theta|y_{obs})$. When $S(y)$ is not sufficient, then there is some information loss, and  $\pi(\theta|s_{obs})$ will be less precise than $\pi(\theta|y_{obs})$. In turn, the ABC posterior approximation $\pi_h^{ABC}(\theta|s_{obs})$ will be less precise than $\pi(\theta|s_{obs})$. While the reduction in dimension from $y_{obs}$ to $s_{obs}$ will allow for increased precision by permitting a reduced kernel scale parameter $h$, this must be offset by any loss of information in the construction of the summary statistics \citep{blum+nps13}.

\begin{table}[hbt]
\begin{tabular}{l}
%\vspace{1mm}
%\rule{\textwidth}{0.1mm}
\underline{ABC importance sampling algorithm}
\vspace{3mm}\\

{\bf Input:} \\An observed dataset, $y_{obs}$. \\A desired number of samples $N>0$.\\ A sampling distribution $g(\theta)$, with $g(\theta)>0$ if $p(\theta)>0$.
 \\A smoothing kernel $K_h$ and scale parameter $h>0$. \\A low-dimensional vector of summary statistics $s=S(y)$.\\ Compute $s_{obs}=S(y_{obs})$.
 \vspace{3mm}
\\

{\bf Iterate:} For $i=1,\ldots,N$:\\
%\begin{enumerate}
%\item
1. Sample a parameter vector from sampling distribution $\theta^{(i)}\sim g(\theta)$.\\
Simulate a dataset from the likelihood given parameter vector $\theta^{(i)}$ as $y^{(i)} \sim L(y | \theta^{(i)})$.\\
Compute the summary statistics $s^{(i)}=S(y^{(i)})$.\\
%\item \label{step2} 
2. Weight each sample $\theta^{(i)}$ by $w^{(i)}\propto K_h(\|s^{(i)}-s_{\text{obs}}\|)p(\theta^{(i)})/g(\theta^{(i)})$.
\vspace{3mm}\\
%\end{enumerate}

{\bf Output:}\\ A set of $i=1,\ldots,N$ weighted samples $(w^{(i)},\theta^{(i)},s^{(i)})$, drawn from $\pi_h^{ABC}(\theta,s|s_{obs})$ (\ref{eqn:ABCjointPostSS}).\\
Or a set of $i=1,\ldots,N$ weighted samples $(w^{(i)},\theta^{(i)})$, drawn from $\pi_h^{ABC}(\theta|s_{obs})$ (\ref{eqn:ABCpostSS}).\\
\hline
%\rule{\textwidth}{0.1mm}
\end{tabular}
\caption{\small A simple ABC importance sampling algorithm, based on a single large sample of size $N$.  \label{alg:ABCimportanceSampling}}
\end{table}

 A precise importance sampling algorithm to generate from $\pi_h^{ABC}(\theta|s_{obs})$ (\ref{eqn:ABCpostSS}) is given by Algorithm \ref{alg:ABCimportanceSampling}.  To see how this works while avoiding evaluation of the intractable likelihood function $L(s|\theta)$, note that this algorithm targets the joint posterior approximation $\pi_h^{ABC}(\theta,s|s_{obs})$ -- marginalising  over $s$ so that the draws come from $\pi_h^{ABC}(\theta|s_{obs})$ is achieved by simply discarding the $s^{(i)}$ in the Monte Carlo output. Hence our sampling distribution must span $(\theta,s)$. In the present setting, it is natural to use
 \[
 	(\theta^{(i)},s^{(i)}) \sim L(s|\theta)g(\theta).
\]
From which the importance weight becomes
\begin{eqnarray*}
	w^{(i)} & \propto &
	\frac{\pi_{h}^{ABC}(\theta^{(i)}|s_{obs})}{L(s^{(i)}|\theta^{(i)})g(\theta^{(i)})}
	\propto
	\frac{K_h(\|s^{(i)}-s_{obs}\|)L(s^{(i)}|\theta^{(i)})p(\theta^{(i)})}{L(s^{(i)}|\theta^{(i)})g(\theta^{(i)})}\\
	& = & \frac{K_h(\|s^{(i)}-s_{obs}\|)p(\theta^{(i)})}{g(\theta^{(i)})},
\end{eqnarray*}
which is conveniently free of computationally intractable likelihood terms.

%%%%%%%%%%%%%%%%
%%%%%%%%%%%%%%%%%
\subsection{A simple example}%%
%%%%%%%%%%%%%%%%%
%%%%%%%%%%%%%%%%
\label{sec:simpleExample}

\begin{figure}[tbp]
  \centering
    \includegraphics[width=0.8\textwidth]{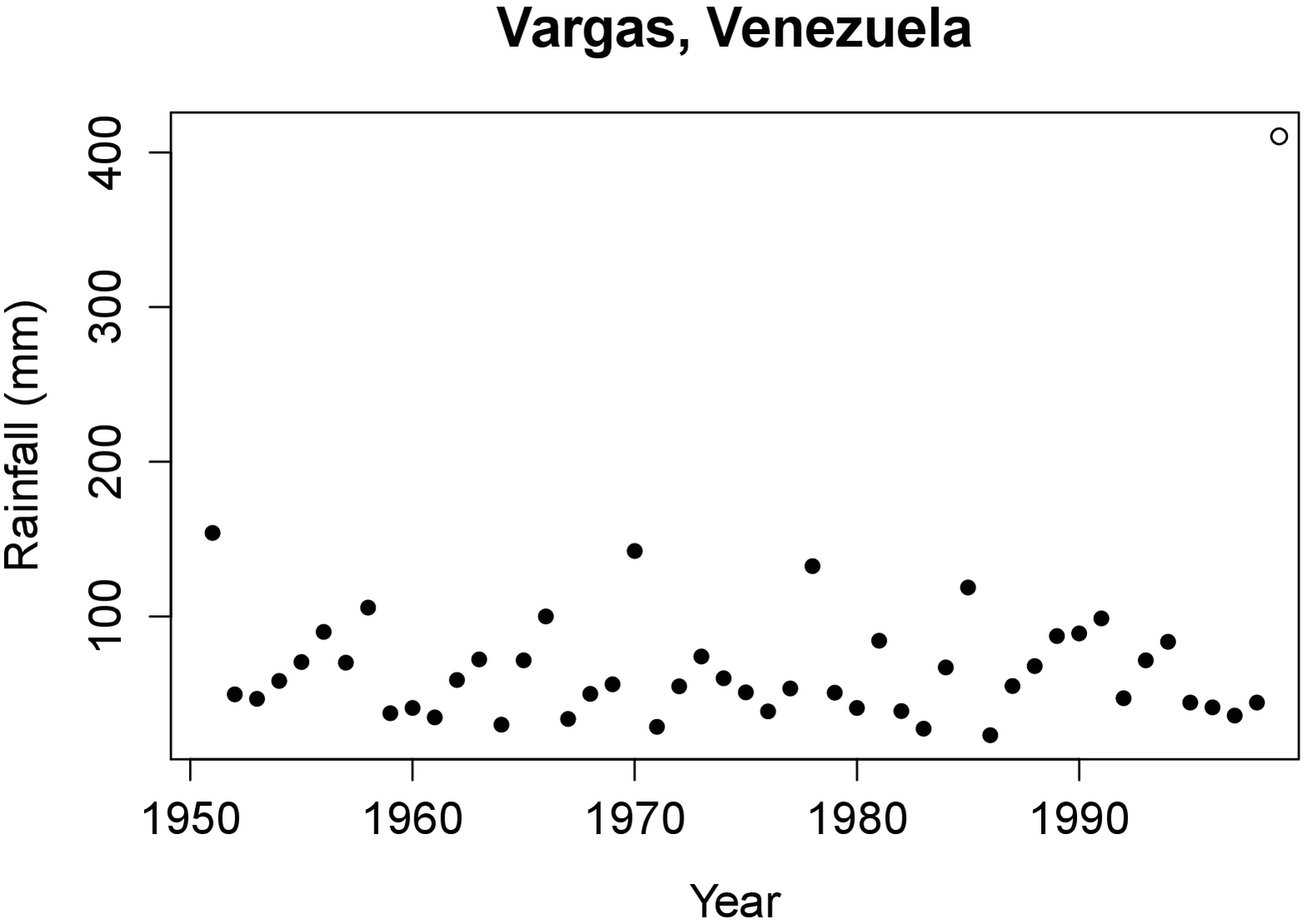}
  \caption
  [Annual maximum daily rainfall (mm) in Vargas.] % short one for ToC.
  {  \label{fig:Figure0}Annual maximum daily rainfall values (mm) 1951--1999 recorded at Maiquetia station on the central coast of Venezuela. The open circle denotes the extreme rainfall observed during the {\it Vargas Tragedy}, in December 1999.}
\end{figure}

As an illustration of ABC methods in a simple setting, consider a standard analysis of univariate block maxima.
Figure \ref{fig:Figure0} illustrates the annual maximum daily rainfall values (measured in millimetres) in the years 1951--1999, recorded at Maiquetia International Airport, on the central coast of Venezuela. This dataset is particularly notable for the unusually extreme daily rainfall events in December 1999 that caused the worst environmentally related tragedy in Venezuelan history, and one of the largest historical rainfall-induced debris flows documented in the world \citep{wieczorek+lemb01}. The resulting human, infrastructure and economic impacts have become known as the {\em Vargas Tragedy}. These data have previously been analysed by \citet{coles+p03} and \citet{coles+ps03}.

%%%%%%%%%%%%%%%%%%%%%%%%%%%%%%%%%%%%%%%%%%%%%%%%%%%
%%%%%%%%%%%%%%%%%%%%%%%%%%%%%%%%%%%%%%%%%%%%%%%%%%%

As these data are univariate annual maxima, they may be approximately modelled by the generalised extreme value (GEV) distribution, which has distribution function shown in (\ref{eq:gev}).
From this, and given a prior specification $p(\mu,\sigma,\xi)$, samples can easily be obtained from the posterior distribution $\pi(\mu,\sigma,\xi|y_{obs})$ using standard posterior simulation algorithms, such as Markov chain Monte Carlo (MCMC).

As the GEV is obviously tractable for analysis, this means that a gold standard result is available to compare with an ABC posterior approximation. Four specifications of the vector of summary statistics are considered for the dataset $y=(y_1,\ldots,y_n)$:
\begin{eqnarray*}
	s_1(y) & = &  (y_1,\ldots,y_n)'=y\\
	s_2(y) & = & (y_{(1)},\ldots,y_{(n)})'\\
	s_3(y) & = & (\hat{\mu}_L,\hat{\sigma}_L,\hat{\xi}_L)'\\
	s_4(y) & = & (\hat{\mu},\hat{\sigma},\hat{\xi})'.
\end{eqnarray*}
Here, $s_1(y)=y$ is the original full dataset, whereas $s_2(y)$ is the vector of $n$ order statistics of $y$ such that $y_{(1)}\leq\ldots\leq y_{(n)}$. The vector $s_3(y)$ consists of the three $L$-moments estimates (following \citet{hosking90}) of each GEV parameter, and $s_4(y)$ is formed from the standard maximum likelihood estimates based on (\ref{eq:gev}). Note that $s_1(y)$, $s_2(y)$ and $s_4(y)$ are sufficient statistics for this analysis, and so $\pi(\mu,\sigma,\xi|s_{obs})=\pi(\mu,\sigma,\xi|y_{obs})$ for these choices as there is no loss of information through the choice of summary statistics. However, both $s_1(y)$ and $s_2(y)$ are high-dimensional ($n=49$), and so in practice, the resulting ABC posterior approximation may suffer from the curse of dimensionality. Both $s_3(y)$ and $s_4(y)$ are low dimensional, with the minimal one summary statistic per model parameter, but $s_3(y)$ is not sufficient for the model parameters, leading to some loss of information. Overall it might be expected that $s_4(y)$ will perform the best, as the dimension of the summary vector is minimised, without any loss of information (it is a minimal sufficient statistic). However, $s_4(y)$ will not typically be available in a typical ABC analysis due to the intractability of the likelihood function.

To complete the Bayesian model specification, the prior is notionally set as $p(\mu,\sigma,\xi)\propto 1$ over the support of the model parameters.
To implement Algorithm \ref{alg:ABCimportanceSampling} a sampling distribution, $g(\mu,\sigma,\xi)$, is required. For this analysis, the prior distribution is improper, and so setting $g(\mu,\sigma,\xi)=p(\mu,\sigma,\xi)$ is not appropriate. In these situations it is common to first identify a region of high posterior density based on a small number of initial simulations of $(\theta^{(i)},y^{(i)})\sim L(y|\theta)b(\theta)$ where $b(\theta)$ is chosen freely by the user. Once the high posterior density region has been identified, then $g(\theta)$ can be specified to be e.g. uniform or proportional to the prior distribution over this region. See e.g. \cite{fearnhead+p12} for an explicit implementation of this idea. In the current setting, we exploit the fact that we know the actual location of the posterior distribution and specify $g(\mu,\sigma,\xi)=U(30,70)\times U(5,45)\times U(-0.3, 1.5)$.

The different scales and correlations  of the summary statistics are taken into account by defining
\begin{equation}
\label{eqn:mahalanobis}
	\|s-s_{obs}\| = \left[(s-s_{obs})'\hat{\Sigma}^{-1}(s-s_{obs})\right]^{1/2},
\end{equation}
as the Mahalanobis distance, where $\hat{\Sigma}$ is an estimate of the covariance of $s_{obs}$. In practice the estimate of the covariance matrix only needs to be approximately correct, and so can be determined by identifying some point in parameter space $(\mu_0,\sigma_0,\xi_0)$ that is likely in an area of high posterior density, generating a number of summary statistic vectors $s^{(i)}\sim L(s|\mu_0,\sigma_0,\xi_0)$ conditional on this point, and then computing the sample covariance matrix of these vectors. See e.g. \citet{luciani+sjft09, sisson+f11} for some examples of this approach. In the present analysis, $(\mu_0,\sigma_0,\xi_0)$ are set as the maximum likelihood estimates $(\hat{\mu},\hat{\sigma},\hat{\xi})$ for convenience.

The following results are based on $N=1,000,000$ samples, where the smoothing kernel $K_h(u)$ is the uniform kernel over $(-h,h)$. The kernel scale parameter is determined as the 0.15, the 0.05 and the 0.005 quantile of $\|s_j^{(i)}-s_{j,obs}\|$ for each of the summary vectors $j=1,\ldots,4$, where $s_{j,obs}=s_j(y_{obs})$. This results in ABC approximations to the posterior distribution constructed from $150,000$, $50,000$ and $5,000$ samples respectively. Note that this retrospective definition of $h$ is not quite in line with Algorithm \ref{alg:ABCimportanceSampling}, which strictly requires $h$ to be determined before the first sample is drawn. However, this procedure of first drawing many $(\theta^{(i)},s^{(i)})$ samples, and then determining $h$ is commonplace in ABC practice (e.g. \citet{beaumont+zb02,blum+nps13}).

%%%%%%%%%%%%%%%%%%%%%%%%%%%%%%%%%%%%%%%%%%%%%%%%%%%
\begin{figure}[tbp]
  \centering
    \includegraphics[width=\textwidth]{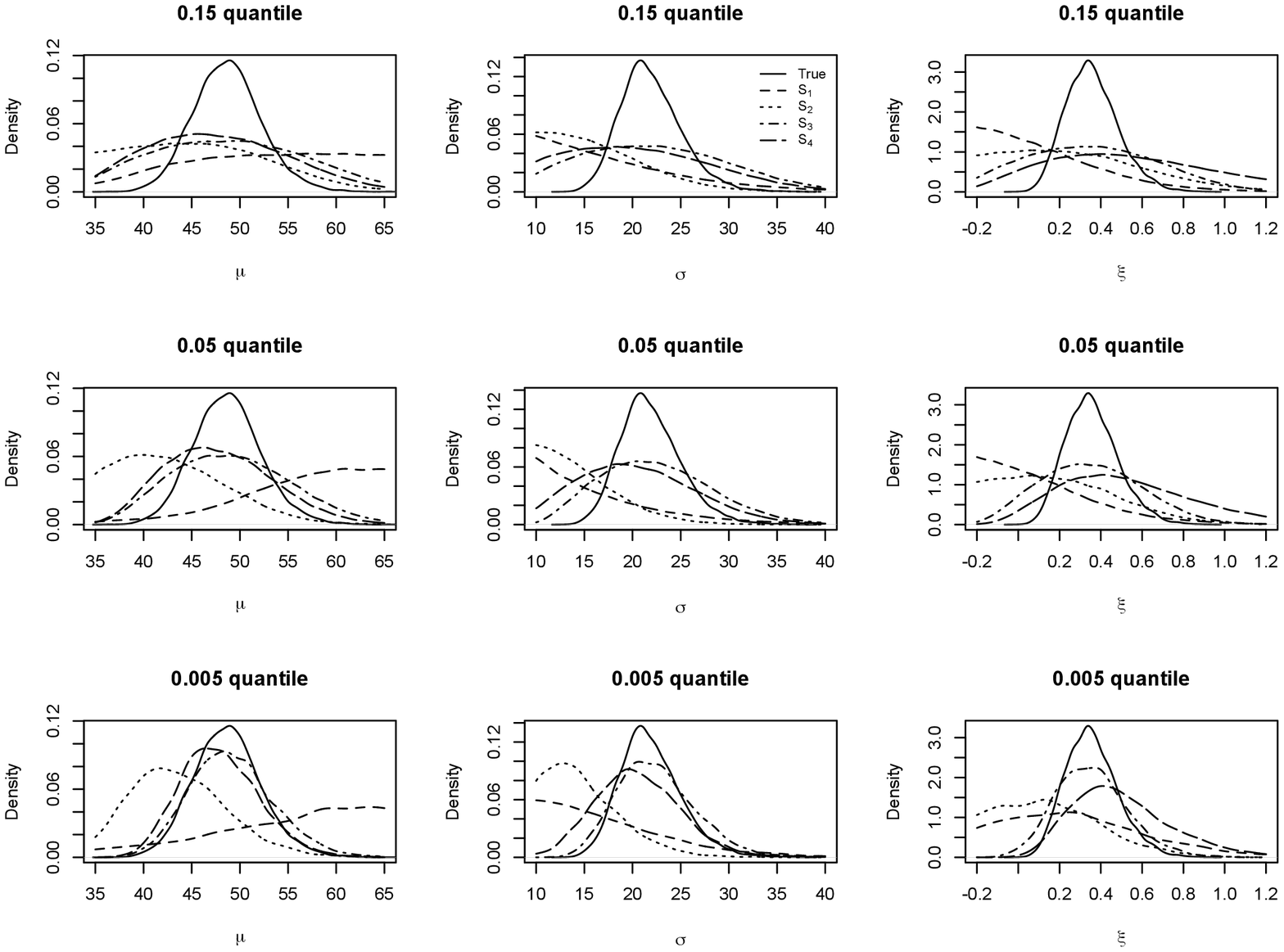}
  \caption
  [ABC approximations of the GEV parameters in the Vargas analysis.] % short one for ToC.
  {  \label{fig:Figure1} Estimates of the marginal posterior distributions of the GEV parameters $\mu$ (left column), $\sigma$ (centre column) and $\xi$ (right column) based on the Vargas dataset. Each panel shows the true posterior marginal distribution (solid line), and the four ABC approximations to the posterior based on the vectors of summary statistics $s_1(y),\ldots,s_4(y)$. Rows indicate the quantile of $\|s_j^{(i)}-s_{j,obs}\|$ used to determine the kernel scale parameter $h$.}
\end{figure}
%%%%%%%%%%%%%%%%%%%%%%%%%%%%%%%%%%%%%%%%%%%%%%%%%%%

Figure \ref{fig:Figure1} illustrates the estimated posterior marginal distributions of $\mu$ (left column), $\sigma$ (centre column) and $\xi$ (right column) for both the true marginal posterior (solid line) and the four ABC marginal posterior approximations (various line types) based on $s_1(y),\ldots,s_4(y)$. Rows correspond to the different values of kernel scale parameter $h$, so that $h$ is decreasing from the top row of Figure \ref{fig:Figure1} to the bottom row.

It is readily apparent that all of the ABC approximations to the posterior distribution based on $s_1(y)=y$ (the dashed line) are very poor at best, regardless of the size of $h$. This should not be a surprise, as $\dim(s_1(y))=49$, and the chance of generating any 49-dimensional vector of independent observations, $s_1^{(i)}$, such that $\|s_1^{(i)}-s_{1,obs}\|\leq h$ becomes vanishingly small as $h$ decreases. Of course, in theory, as $h\rightarrow 0$ then the resulting ABC posterior approximation will be equivalent to the true posterior distribution $\pi(\mu,\sigma,\xi|y_{obs})$ because $s_1(y)$ is a sufficient statistic. However, even taking the closest 5,000 samples (Figure \ref{fig:Figure1}, bottom row with the 0.005 quantile) out of 1 million was not enough to produce any reasonable accuracy in this case. This means that for practical purposes, it is not viable to use $s_1(y)$.

The ABC posterior approximations based on $s_2(y)$ (dotted lines) are as poor as those based on $s_1(y)$ for high values of $h$. However, there is some evidence that the density estimates are beginning to improve when $h$ is reduced to the 0.005 quantile of $\|s_2^{(i)}-s_{2,obs}\|$. This is more obvious for the location parameter $\mu$, which is typically the easiest parameter to estimate in any analysis, although there is little change for $\sigma$ and $\xi$. The reason why the 49-dimensional vector $s_2(y)$ can achieve better performance than $s_1(y)$ (also in 49 dimensions) is that a close match of $s_2^{(i)}$ to $s_{2,obs}$ is far more likely to occur due to the induced dependence of the summary statistics. For example, consider that (say) the minimum values of each of $y$ and $y_{obs}$ are more likely to be close than a randomly chosen element from each dataset, which is the comparison being made when using $s_1(y)$. However, as with $s_1(y)$, the vector of summary statistics $s_2(y)$ still does not appear to be practically viable with a computational overhead of only $N=1,000,000$ samples.

However, the vectors of 3 dimensional summary statistics $s_3(y)$ (dot-dash line) and $s_4(y)$ (long dashed line) both produce more practically useful ABC posterior approximations. The broad adherence of the approximate densities to the true posterior marginal distributions is apparent in Figure \ref{fig:Figure1} even at the 0.15 quantile, and their accuracy only increases as $h$ decreases. At the 0.005 quantile, either posterior is almost good enough to use in practice,
although the $L$-moments based ABC approximation does appear to slightly better approximate the true posterior in this case, despite the information loss in $s_3(y)$.
It seems quite likely that increasing $N$, thereby allowing $h$ to be decreased further, would result in a viable ABC posterior approximation based on $s_3(y)$. While in theory using $s_4(y)$ should be slightly more efficient than using $s_3(y)$, this choice of summary statistics will not be available in practice.

In Section \ref{sec:OtherABC}  a method is discussed -- the regression adjustment -- that will allow for an improved accuracy in an ABC analysis, without the need to increase computational overheads ($N$) further.
For a technique to determine whether the chosen value of $h$ is ``low enough'' for a good ABC posterior approximation when the true posterior distribution is unknown, see \cite{prangle+bps14}.

Overall, the main concepts of ABC have been illustrated by this simple analysis. ABC methods are themselves a simple and intuitive procedure that can produce viable approximations to posterior distribution without the need to evaluate the likelihood function. They work best when the vector of summary statistics are both highly informative for the model parameters, and are low dimensional. However, even then, ABC methods can have large computational overheads if high accuracy is desired. This is the price to pay for not being able to evaluate the likelihood function.

%%%%%%%%%%%%%%%%%%%%
%%%%%%%%%%%%%%%%%%%%%
\subsection{Other useful ABC methods}%%
%%%%%%%%%%%%%%%%%%%%%
%%%%%%%%%%%%%%%%%%%%
\label{sec:OtherABC}

%%%%%%%%%%%%%%%%%%%%%%%%%%
%%%%%%%%%%%%%%%%%%%%%%%%%%%
\subsubsection{Regression-adjustment techniques}%%
%%%%%%%%%%%%%%%%%%%%%%%%%%%
%%%%%%%%%%%%%%%%%%%%%%%%%%

As previously discussed, standard ABC methods suffer from the curse of dimensionality \citep{blum10}, so that the kernel scale parameter $h$ must increase for practical purposes as the number of summary statistics increases. As a result, there is still often a large discrepancy between $s^{(i)}$ and $s_{obs}$ in the final sample from $\pi_h^{ABC}(\theta,s|s_{obs})$, whereas greatest accuracy of the ABC posterior approximation is obtained if $s^{(i)}\approx s_{obs}$ (i.e. if $h\rightarrow 0$). Regression-adjustment techniques aim to reduce this discrepancy by explicitly modelling the relationship between the sampled $\theta^{(i)}$ and $s^{(i)}$, and then adjusting $(\theta^{(i)},s^{(i)})\rightarrow(\theta^{(i)*},s^{(i)*})$ so that $s^{(i)*}=s_{obs}$. The resulting (weighted) samples $\theta^{(i)*}$ will then form an improved approximation to $\pi(\theta|s_{obs})$ than those (weighted) $\theta^{(i)}$ from $\pi_h^{ABC}(\theta|s_{obs})$, if the regression model is correct.

The simplest form of a model for this is a homoscedastic regression in the region of $s_{obs}$, so that
\[
	\theta^{(i)}=m(s^{(i)}) + e^{(i)},
\]
where $m(s^{(i)})=\mathbb{E}[\theta|s=s^{(i)}]$ is the mean function, and the $e^{(i)}$ are zero mean, common variance random variates.  In the case of a local-linear regression we have the model
\[
	m(s^{(i)}) = \alpha + \beta's^{(i)}.
\]
\citet{beaumont+zb02} estimated this model by minimising the least squares criterion $\sum_{i=1}^N\omega^{(i)}\|m(s^{(i)})-\theta^{(i)}\|^2$ where $\omega^{(i)}=K_h(\|s^{(i)}-s_{obs}\|)$. The regression-adjusted weighted sample $(\theta^{(i)*},w^{(i)})$ drawn approximately from $\pi(\theta|s_{obs})$ is then obtained as
\begin{equation}
\label{eqn:regAdjust}
	\theta^{(i)*} = \hat{m}(s_{obs})+(\theta^{(i)}-\hat{m}(s^{(i)})),
\end{equation}
where $\hat{m}(s)=\hat{\alpha}+\hat{\beta}'s$ is the  fitted mean function. In addition to the local-linear regression-adjustment of \citet{beaumont+zb02}, variations on this approach include a non-linear, heteroscedastic regression-adjustment \citep{blum+f10} and a ridge regression-adjustment \citep{blum+nps13}. A qualitatively different, but also useful form of marginal adjustment to improve ABC posterior approximation accuracy, that can be used in conjunction with the regression-adjustment is described by \citet{nott+fms14}.

%%%%%%%%%%%%%%%%%%%%%%%%%%%%%%%%%%%%%%%%%%%%%%%%%%%
\begin{figure}[tbp]
  \centering
    \includegraphics[width=\textwidth]{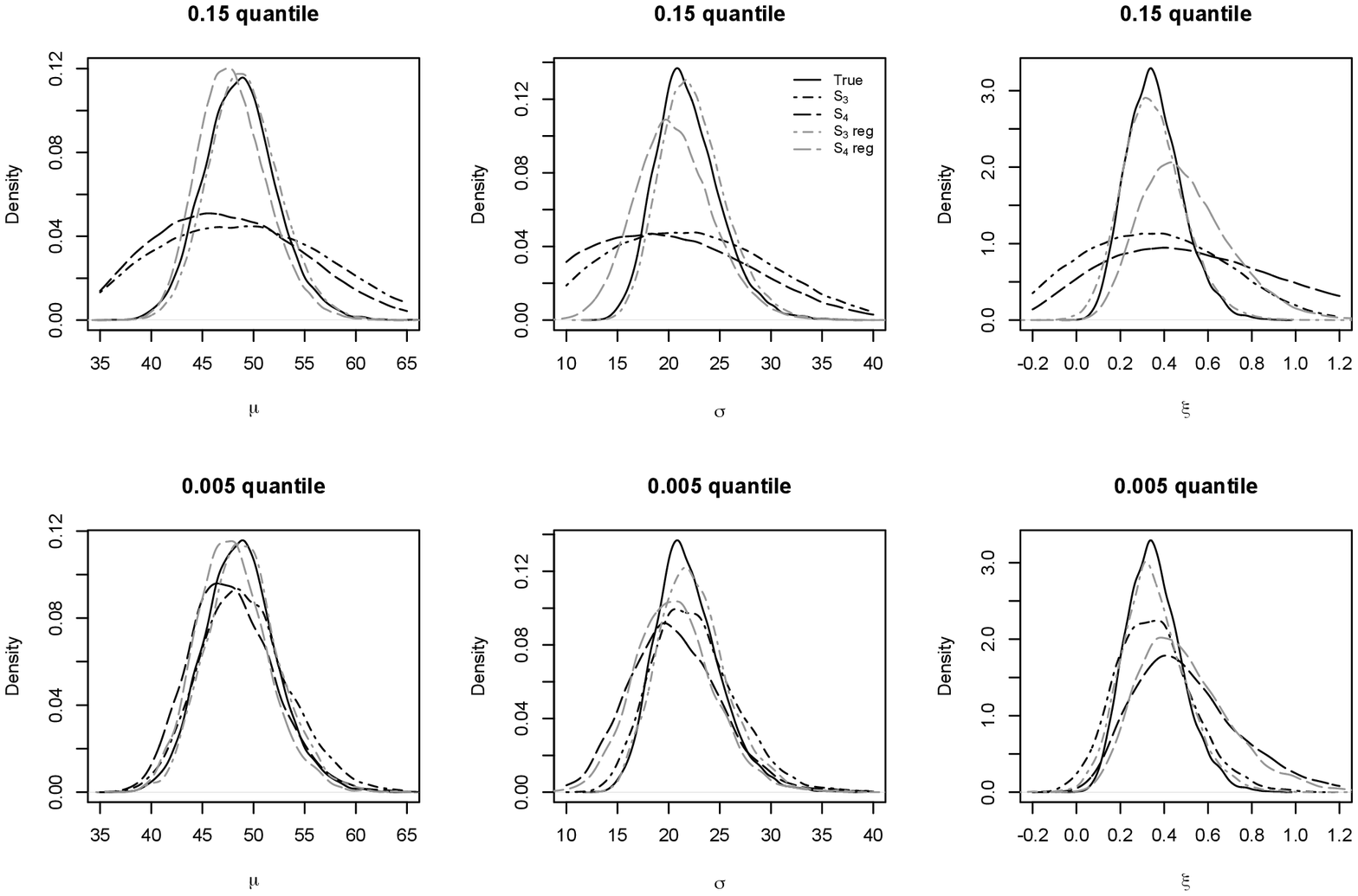}
  \caption
  [Regression-adjusted ABC approximations of the GEV parameters in the Vargas analysis.] % short one for ToC.
  {\label{fig:Figure2}Regression-adjusted estimates of the marginal posterior distributions of the GEV parameters $\mu$, $\sigma$ and $\xi$ based on the Vargas dataset. Each panel shows the true posterior marginal distribution (solid line) and the two ABC approximations to the posterior based on the vectors of summary statistics $s_3(y)$ and $s_4(y)$. ABC approximations without regression adjustment are illustrated with black lines. Those with a subsequent regression-adjustment are shown with grey lines.}
\end{figure}
%%%%%%%%%%%%%%%%%%%%%%%%%%%%%%%%%%%%%%%%%%%%%%%%%%%

Figure \ref{fig:Figure2} illustrates the ABC approximations to the posterior marginal distributions of the GEV analysis described in Section \ref{sec:simpleExample}, focusing on the 3-dimensional vectors of summary statistics $s_3(y)$ and $s_4(y)$. The black lines indicate the original approximations shown in Figure \ref{fig:Figure1}, whereas the grey versions of these lines show the same samples following the local-linear regression adjustment (\ref{eqn:regAdjust}).

In all cases, the regression-adjustment has made a noticeable improvement in the quality of the ABC posterior approximation, and in some cases the resulting density estimate is indistinguishable from the truth. This is a strong indicator that regression-adjustment methods should be routinely used in ABC analyses.  It is interesting to note that the regression adjustment produces effectively the same approximation whether the initial samples $(\theta^{(i)},s^{(i)})$ are obtained using relatively high (top row) or low (bottom row) values of kernel scale parameter $h$. This will occur when the fitted regression model remains effectively unchanged as $h$ is varied. In this case, large computational savings can occur by obtaining an initial ABC approximation based on a larger value $h$, and then performing the adjustment, rather than doing the same but with a lower value of $h$.

%%%%%%%%%%%%%%%%%%%%%%
%%%%%%%%%%%%%%%%%%%%%%%
\subsubsection{Choice of summary statistics}%
%%%%%%%%%%%%%%%%%%%%%%%
%%%%%%%%%%%%%%%%%%%%%%

Summary statistic identification is essentially a problem of dimension-reduction with respect to sufficiency for the target model. This is a huge research area in its own right. \citet{blum+nps13} provide a recent comprehensive and comparative review of the methods that have been proposed for dimension reduction in ABC. These can be classified as best subset selection methods,
projection techniques,
regularisation approaches
and other principled methods such as those based on indirect inference.

Some more recent developed approaches  include those by \citet{drovandi+pl14,ruli+sv13} in addition to those particular cases where ABC can be reliably performed using the full original dataset, e.g. due to a likelihood factorisation \citep{barthelme+c14,bonassi+yw11,fan+ns13, jasra+smm12}.

While improved techniques in this area are still being developed, it is probably accurate to say that there is no single ``best'' method. The optimum approach to identifying good summary statistics currently remains careful consideration of the specific model and analysis at hand.

%%%%%%%%%%%%%%%%%%%
%%%%%%%%%%%%%%%%%%%%
\subsubsection{Other ABC algorithms}%%
%%%%%%%%%%%%%%%%%%%%
%%%%%%%%%%%%%%%%%%%
\label{sec:otherABC}

The ABC method listed in Algorithm \ref{alg:ABCimportanceSampling} is based on importance sampling. However there are many other standard algorithms for generating draws from posterior distributions, such as those based on MCMC and sequential Monte Carlo. Each of these approaches are beneficial in particular circumstances \citep{brooks+gjm11,doucet+fg01}. A variety of ABC versions of these algorithms have been developed, and the basic mechanism by which they work is essentially the same as the importance sampling algorithm.

Essentially, each posterior simulation algorithm requires the computation of either importance weights, which are the ratio of the target distribution to the sampling distribution $\pi(\theta|y_{obs})/g(\theta)$, or MCMC acceptance probabilities which are constructed from the ratio of these,  as
\[
	\pi(\theta'|y_{obs})g(\theta|\theta')/\left[\pi(\theta|y_{obs})g(\theta'|\theta)\right],
\]
where the sampling distribution $g$ is now a Markov proposal distribution. In the ABC framework, the approximate posterior distribution is augmented by the auxiliary datasets so that the target distribution is $\pi_h^{ABC}(\theta,s|s_{obs})\propto K_h(\|s-s_{obs}\|)L(s|\theta)p(\theta)$ (\ref{eqn:ABCjointPostSS}). This means that in any posterior simulation algorithm, the sampling or proposal distribution must also be defined on the same space, so that $q(\theta,s)\propto L(s|\theta)g(\theta)$ or $q(\theta,s|\theta',s')\propto L(s|\theta)g(\theta|\theta')$. As a result, when computing an importance weight or acceptance probability targeting $\pi_h^{ABC}(\theta,s|s_{obs})$, the intractable likelihood term $L(s|\theta)$ will always cancel out, as demonstrated at the end of Section \ref{sec:ABCbasics}. This  leaves only computationally tractable terms, which can then be evaluated. \citet{sisson+f11} demonstrate this in detail for the Metropolis-Hastings MCMC algorithm.

For further information on MCMC algorithms for ABC see e.g. \citet{marjoram+mpt03,bortot+cs07,sisson+f11}, and for sequential Monte Carlo algorithms refer to e.g. \citet{sisson+ft07,beaumont+mcr09,drovandi+p11,delmoral+dj12}.

Alternative approaches to constructing an ABC approximation to the posterior distribution, based on building structured density estimates of the sample $(\theta^{(1)},s^{(i)}),\ldots,(\theta^{(N)},s^{(N)})$, have been developed by \citet{bonassi+yw11} and \citet{fan+ns13}.

%%%%%%%%%%%%%%%%%%%%
%%%%%%%%%%%%%%%%%%%%%
\section{ABC for stereological extremes}%%
%%%%%%%%%%%%%%%%%%%%%
%%%%%%%%%%%%%%%%%%%%
\label{sec:stereo}

In the production of clean steels, microscopically small particles termed {\it inclusions} are introduced during the production process. Metallurgic considerations indicate that the strength of the steel is directly related to the size of the largest inclusion in the block, and so inference on the largest inclusion size is important. Commonly, the sampling of inclusions involves measuring the maximum cross-sectional slice of each observed inclusion, $y_{obs}=(y_{1,obs}, \ldots, y_{n,obs})$, obtained from a two-dimensional planar slice through the steel block. Each cross-sectional inclusion size is greater than some measurement threshold, $y_{i,obs}>u$.
The inferential problem is to analyse the unobserved distribution of the largest inclusion in the block, based on the information in the cross-sectional slice, $y_{obs}$. The focus on the size of the largest inclusion means that this is an extreme value variation on the standard stereological problem \citep{baddeley+j05}.

\citet{anderson+c02} proposed a mathematical model for those observed cross-sectional measurements.
  The proposed model assumed that the inclusions were spherical with diameters $V$, that their centres followed a homogeneous Poisson process with rate $\lambda>0$ in the volume of steel, and that the inclusion diameters were mutually independent and independent of inclusion location. The distribution of the largest inclusion diameters, $V|V>v_0$,  (i.e. those conditional on exceeding some threshold, $v_0$) was assumed to be well approximated by a generalised Pareto distribution, with distribution function
\begin{equation}
\label{eqn:gpd}
	\mbox{Pr}(V\leq v|V>v_0) = 1-\left[1+\frac{\xi(v-v_0)}{\sigma}\right]^{-1/\xi}_+,
\end{equation}
for $v>v_0$, where $[a]_+=\max\{0,a\}$, and $\sigma>0$ and $-\infty<\xi<\infty$ are scale and shape parameters, following standard extreme value theory arguments \citep{coles01}. Accordingly the parameters of the full spherical inclusion model are  $\theta=(\lambda, \sigma, \xi)$.

Each observed cross-sectional inclusion diameter, $y_{i,obs}$, is associated with an unobserved true inclusion diameter $V_i$.
However, note that the probability of observing the cross-sectional diameter size $y_{i,obs}$ (where $y_{i,obs}\leq V_i$) is dependent on the value of $V_i$, as larger inclusion diameters give a greater chance that the inclusion will be observed in the two-dimensional planar cross-section. The number of observed inclusions, $n$, is also a random variable. In these terms, interest is in the distribution of the largest inclusion diameters, $V_1,\ldots,V_n$, given the observed cross-sectional measurements, $y_{1,obs},\ldots,y_{n,obs}$.

\cite{anderson+c02} were able to construct a likelihood function for this model by adapting the solution to Wicksell's corpuscle problem \citep{wicksell25}. They also overcome numerical difficulties with the resulting likelihood function  of the model by treating the unobserved $V_i$ as latent variables in a Bayesian hierarchical formulation. However, while their model assumptions of a Poisson process and inclusion independence are not unreasonable, the assumption that the inclusions are spherical is not plausible in practice.

Because of this,
a generalisation of the spherical inclusion model was proposed by \cite{bortot+cs07}, who considered a family of ellipsoidal inclusions. In addition to the previous Poisson process and independence assumptions, this new model considered inclusions to be ellipsoidal and randomly oriented in space, with principal diameters $(V^1,V^2,V^3)$.  As before, the distribution function of $V^3|V^3>v_0$ is specified as the generalised Pareto distribution (\ref{eqn:gpd}), where $V^3$ is defined as the largest ellipsoidal diameter. The other two principal diameters are defined as
$V^j=U_jV^3$ for $j=1,2$, where $U_1$ and $U_2$ are independent uniform $U(0,1)$ variables. Finally, in order to avoid ambiguity, the observed cross-sectional measurement $s_{i,obs}$ is  assumed to be the largest principal diameter of the ellipse generated by the planar section of an ellipsoidal inclusion.

While the ellipsoidal inclusion is more realistic than the spherical inclusion model, there are analytic and computational difficulties in extending likelihood-based inference to more general families of inclusion \citep{baddeley+j05,bortot+cs07}. As a result ABC methods are a good candidate procedure to approximate the posterior distribution $\pi(\lambda,\sigma,\xi|y_{obs})$ under the ellipsoidal model.

In the following analysis, the prior $p(\lambda,\sigma,\xi)\propto 1$ is specified as uniform over the support of the model parameters. For the purposes of implementing Algorithm \ref{alg:ABCimportanceSampling} total of $N=2$ million samples were drawn from the sampling distribution $g(\lambda,\sigma,\xi)=U(10,80)\times U(0,10)\times U(-3,3)$ for the spherical model and $g(\lambda,\sigma,\xi)=U(60,130)\times U(0,10)\times U(-3,3)$ for the ellipsoidal model based on a pilot analysis.
A 7-dimensional vector of  summary statistics was specified as
\begin{equation}
\label{eqn:bortotSS}
	S(y)=(q_{0.5}(y),q_{0.7}(y),q_{0.9}(y),q_{0.95}(y),q_{0.99}(y),q_{1}(y),n')
\end{equation}
where $q_a(y)$ denotes the $a$-th quantile of the dataset $y$ (with interpolation if necessary), and $n'$ is the number of observations in $y$. The Mahalanobis distance (\ref{eqn:mahalanobis}) is used to compare auxiliary and observed summary statistics $\|s-s_{obs}\|$, where the covariance matrix $\hat{\Sigma}$ is either the identity matrix $I$, or an estimate of $\mbox{Cov}(s|\lambda_0,\sigma_0,\xi_0)$ based on 1,000 samples $s\sim L(s|\lambda_0,\sigma_0,\xi_0)$, where $(\lambda_0,\sigma_0,\xi_0)=(30,1.5,-0.05)$ for the spherical model and $(\lambda_0,\sigma_0,\xi_0)=(95,1.9,-0.1)$ for the ellipsoidal model. The values for $(\lambda_0,\sigma_0,\xi_0)$ were based on preliminary analyses for each model. The smoothing kernel $K_h(u)$ is the uniform kernel over $(-h,h)$ and the kernel scale parameter $h$ is determined as the 0.001 quantile of the $N$ samples $\|s^{(i)}-s_{obs}\|$, resulting in 2,000 samples from the approximate posterior $\pi_h^{ABC}(\lambda,\sigma,\xi|s_{obs})$.
The observed dataset $s_{obs}$ is derived from a set of 112 inclusion diameters previously analysed by \citet{anderson+c02} and \citet{bortot+cs07}

%%%%%%%%%%%%%%%%%%%%%%%%%%%%%%%%%%%%%%%%%%%%%%%%%%%
\begin{figure}[tbp]
  \centering
    \includegraphics[width=\textwidth]{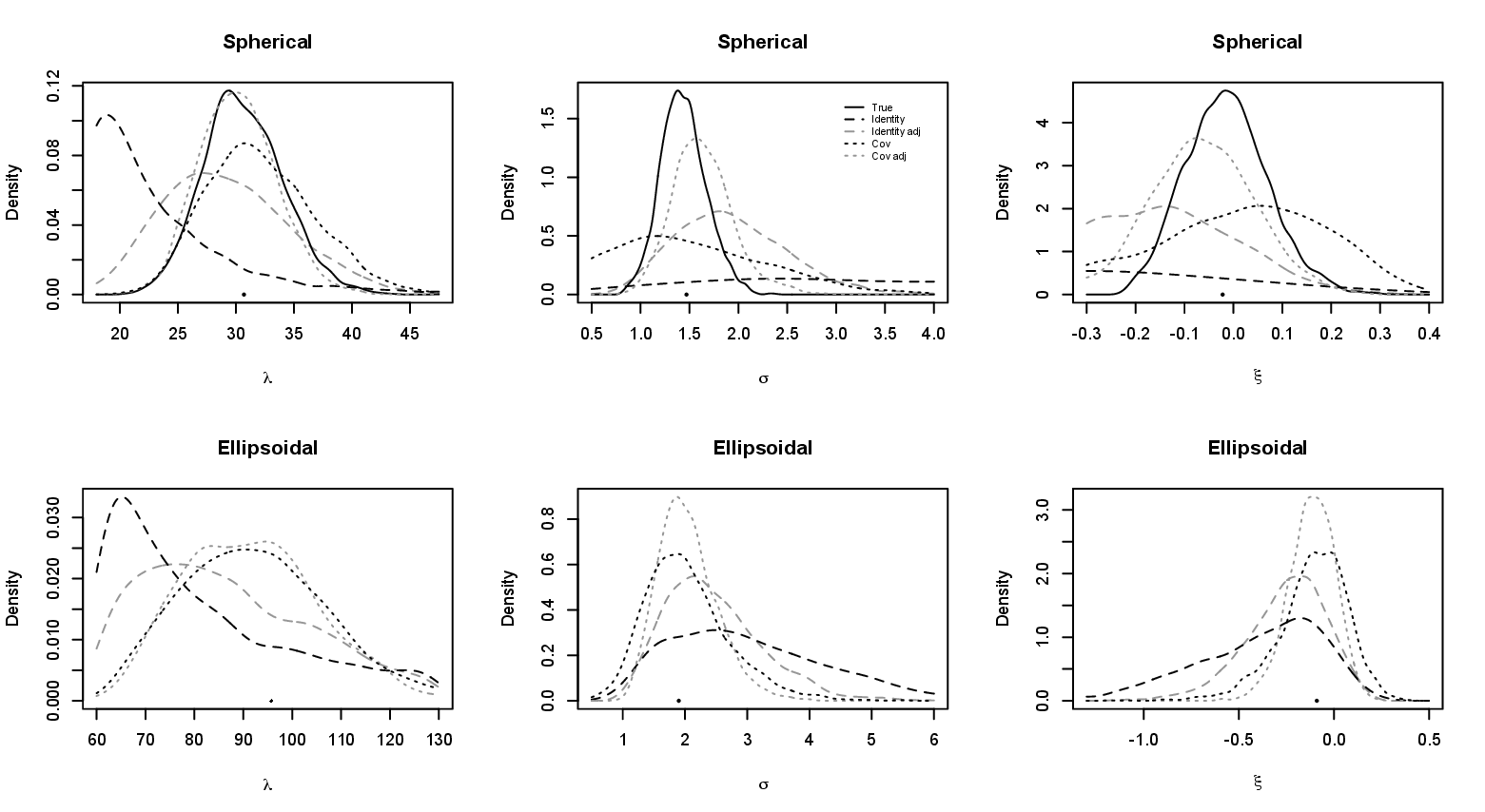}
  \caption
  [Marginal posterior estimates of $\lambda$, $\sigma$ and $\xi$ for the stereological extremes analysis] % short one for ToC.
  {\label{fig:Bortot1}Marginal posterior estimates of $\lambda$, $\sigma$ and $\xi$ for the stereological extremes analysis. Top and bottom rows respectively correspond to the spherical and ellipsoidal inclusion models. Each panel shows the true posterior marginal distribution (solid line -- spherical model only), and the ABC approximations to the posterior based on the summary statistics (\ref{eqn:bortotSS}). Dashed lines indicate ABC posterior approximation using the identity matrix $\hat{\Sigma}=I$ in the Mahalanobis distance, dotted lines use an estimate of $\hat{\Sigma}=\mbox{Cov}(s|\lambda_0,\sigma_0,\xi_0)$. Grey versions of each line indicate a subsequent regression-adjusted estimate. Points show the marginal posterior means estimated by \citet{bortot+cs07}.}
\end{figure}
%%%%%%%%%%%%%%%%%%%%%%%%%%%%%%%%%%%%%%%%%%%%%%%%%%%

Figure \ref{fig:Bortot1} illustrates various marginal posterior estimates of $\pi(\lambda,\sigma,\xi|s_{obs})$ for both the spherical (top row) and ellipsoidal (bottom row) inclusion models. The ABC approximation to the posterior $\pi_h^{ABC}(\lambda,\sigma,\xi|s_{obs})$ using the identity matrix $\hat{\Sigma}=I$ is illustrated by the dashed lines, and the approximation using $\hat{\Sigma}=\mbox{Cov}(s|\lambda_0,\sigma_0,\xi_0)$ is shown by the dotted lines. Grey lines denote the same approximations but with a subsequent regression adjustment. The solid black line indicates an estimate of the true posterior $\pi(\lambda,\sigma,\xi|y_{obs})$ following the latent variable model of \citet{anderson+c02}, for the spherical inclusion model only. The dots show the marginal posterior means estimated by \citet{bortot+cs07}.

For the spherical inclusions model (Figure \ref{fig:Bortot1}, top row), there is a clear difference in the resulting estimates of the marginal posterior distributions due to the choice of covariance matrix in the comparison $\|s^{(i)}-s_{obs}\|$. When $\hat{\Sigma}$ approximates $\mbox{Cov}(s|\lambda_0,\sigma_0,\xi_0)$, the ABC approximation (dotted lines) is able to locate the true posterior density (solid line) fairly well. However, when the identity matrix is used (dashed lines), the approximation is substantially worse. The primary reason for this is that the summary statistics with the highest variability -- namely $n'$ and the highest quantiles of $y$ -- dominate the comparison, so $\|s^{(i)}-s_{obs}\|$ is likely to be large unless these highly variable statistics happen to be close to those in $s_{obs}$. When summary statistic scaling (and correlation) is taken into consideration, the relative closeness of $s^{(i)}$ to $s_{obs}$ can be better measured.

In addition,  the implementation of a regression-adjustment also makes a substantial improvement in the quality of the ABC approximation, even for the very poor estimates using $\hat{\Sigma}=I$. The best results are clearly obtained when $\hat{\Sigma}\approx\mbox{Cov}(s|\lambda_0,\sigma_0,\xi_0)$ is used, followed by a regression adjustment. Indeed, the inclusion rate parameter $\lambda$ is almost perfectly estimated marginally. More detailed analysis (not shown) suggests that the results may improve further if the kernel scale parameter $h$ could be lowered further. However, this would require a larger number of initial simulations, $N$.

For the ellipsoidal inclusions model (Figure \ref{fig:Bortot1}, bottom row), while estimates of the true marginal posterior densities of $\pi(\lambda,\theta,\xi|y_{obs})$ are not available, qualitatively similar conclusions to the spherical inclusions model can be made. The best performing ABC approximation is likely to be when accounting for the scale and correlation of the summary statistics, followed by a regression adjustment.
This approximation agrees well with the previous estimates of marginal posterior means for these data given by \citet{bortot+cs07}

A primary difference between the parameter estimates from the two models is that the number (rate) of inclusions is much higher for the ellipsoidal model. This reflects that
the overall dimensions of an ellipsoid are smaller than a sphere with diameter the same as the largest principal diameter of the ellipsoid. Hence, given the observed planar intersections, a smaller rate of inclusion is predicted under the spherical inclusion model than under the ellipsoidal inclusion model.
This analysis suggests that measures of inclusion impact that depend strongly on the rate of extreme inclusions are likely to be strongly sensitive to assumptions on inclusion shape.

While it is possible to perform Bayesian model selection through the computation of Bayes Factors in the ABC framework \citep{robert+mp11,marin+pnrr14}, in the present setting it is doubtful if the current data -- the maximum cross-sectional slice of each observed inclusion -- is informative for this quantity. To proceed further along the road of model choice, further measurements on each cross-sectional inclusion, such as the minimum and maximum diameters, would be required.

%%%%%%%%%%%%%%%%%%%%
%%%%%%%%%%%%%%%%%%%%%
\section{ABC and max-stable processes}%
%%%%%%%%%%%%%%%%%%%%%
%%%%%%%%%%%%%%%%%%%%
\label{sec:maxstab}
\subsection{Background}
Suppose that there is interest in using ABC methods to fit max-stable processes to extreme temperature data in the context of actuarial risk estimation for a class of financial products known as weather derivatives.  Weather derivatives are contracts that specify a weather reporting station, a time period, and payments corresponding to certain pre-determined weather events.  The intention is for the buyer of this contract to be compensated by the seller if certain undesirable weather events occur (low snowfall for ski resorts, low rainfall for farms, high temperatures for electricity producers, etc.), but in practice, weather derivatives can also be bought and sold for purely speculative reasons.

Denoting the weather random variable by $Y$, then the resulting payment is also a random variable $P(Y)$. Here we are interested in derivatives with payments triggered by events $\{\max Y \ge u\}$ or $\{\min Y \le u\}$ where $Y$ is the variable of interest and $u$ is some threshold meant to target extremes.  For  example, a payment $P$ can be triggered if the maximum temperature in July exceeds 100 degrees Fahrenheit, with payment \$1000 for each degree in excess. This is shown mathematically as
\[
P(Y) = \max \left(1000 \cdot \left(\max_{Y \in July}(Y) - 100\right), 0\right).
\]
Because weather random variables $Y$ can be positively correlated due to close proximity, financial outcomes of weather derivatives can also be positively correlated.  When one considers the total payment from a collection of $D$ weather derivatives $P = P_1 + \ldots + P_D$, recognition of the dependence is essential for fully characterising the distribution of $P$.  The use of max-stable processes is one way to incorporate both spatial dependence and target extremes.

Max-stable processes are the infinite dimensional generalization of multivariate extreme value theory.  Let $Z(x), x \in X \subseteq \mathbb{R}^{p}$ be a spatial process.  If for all $n \ge 1$, there exists sequences $a_{n}(x), b_{n}(x)$, $x \in X$ such that for any $x_{1}, ..., x_{D} \in X$,
\begin{equation*}
P^{n} \left(\frac{Z(x_{d}) - b_{n}(x_{d})}{a_{n}(x_{d})} \leq z(x_{d}), d = 1, ..., D \right) \rightarrow G_{x_{1}, ..., x_{D}}(z(x_{1}), ..., z(x_{D}))
\end{equation*}
then $G_{x_{1}, ..., x_{D}}$ is a multivariate extreme value distribution.  If the above holds for all possible subsets $x_{1}, ..., x_{D} \in X$ for any $D \ge 1$, then the process is max-stable.

A method for constructing such processes was given by de Haan \citep{de1984spectral, deextreme}.  Let $Y(x)$ be a non-negative stationary process on $\mathbb{R}^{p}$ such that $\mathbb{E}(Y(x)) = 1$ at each $x$.  Let $\Pi$ be a Poisson process on $\mathbb{R}_{+}$ with intensity $s^{-2}ds$.  If $Y_{i}(x)$ are independent replicates of $Y(x)$, then
\begin{equation*}
Z(x) = \max s_{i} \cdot Y_{i}(x), \hspace{5mm} x \in X
\end{equation*}
is a stationary max-stable process with unit Fr\'{e}chet margins .  Different choices of the process $Y(x)$ give different max-stable processes.  \citet{smith1990max} described a convenient interpretation of this process, with $\mathbb{R}^{p}$ as the space of storm centers, $s_{i}$ as the magnitude of the $i^{th}$ storm, and $Y_{i}(x)$ as the shape of the $i^{th}$ storm.  The maximum of independent storms is taken to be the max-stable process.

\begin{figure}
\begin{center}
\includegraphics[width=3.5in]{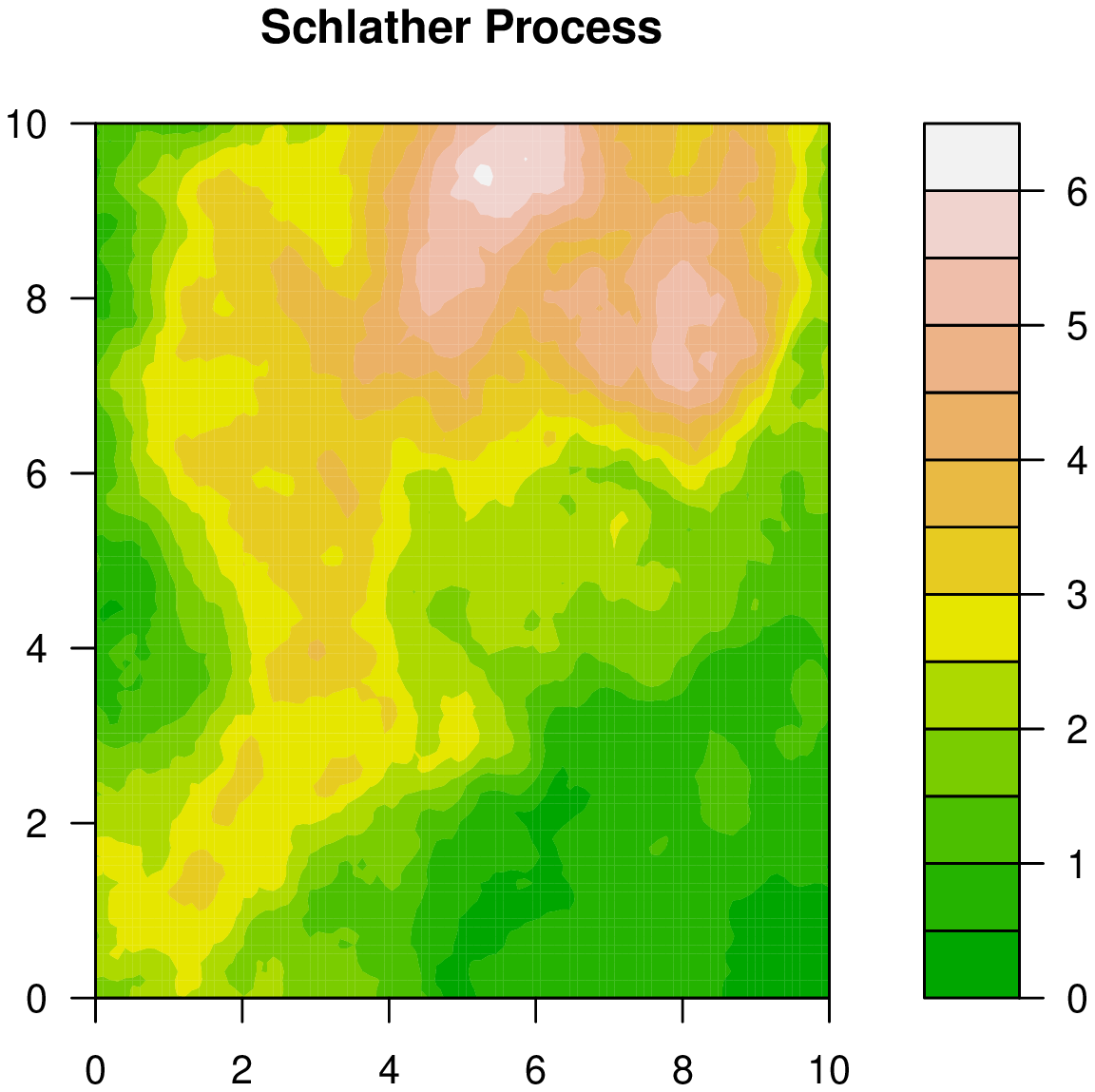}
\caption[Extremal Gaussian process.]{Extremal Gaussian process with Whittle-Mat\'{e}rn correlation with nugget $c_1=1$, range $c_2=3$, and smooth $\nu=1$.}
\label{fig:schlather1}
\end{center}
\end{figure}

\citet{schlather2002models} extended $Y(x)$ to be any stationary Gaussian process on $\mathbb{R}^{p}$ with correlation function $\rho(\cdot)$ and finite mean $\mu = \mathbb{E}[ \max(0, Y(x))] \in (0, \infty)$.  With $s_{i}$ as a Poisson process on $(0, \infty)$ with intensity measure $\mu^{-1}s^{-2}ds$, the quantity
\begin{equation*}
Z(x) = \max_{i} s_{i} \max(0, Y_{i}(x))
\end{equation*}
is a stationary max-stable process with unit-Fr\'{e}chet margins.  The bivariate distribution function is
\begin{equation}
P(Z_{1} \leq z_1, Z_{2} \leq z_2) = \exp \left[- \frac{1}{2} \left(\frac{1}{z_1} + \frac{1}{z_2} \right) \left(1 + \sqrt{1 - 2(\rho(h)+1)\frac{z_1 z_2}{(z_1+z_2)^{2}}} \right) \right]
\label{eq:schlather2}
\end{equation}
where $\rho(h)$ is the correlation of the underlying Gaussian process $Y$ and $h=||x_{1}-x_{2}||$.  The correlation function may be chosen from one of the valid families of correlations for Gaussian processes, with one common choice being the Whittle-Mat\'{e}rn correlation,
\begin{equation}
\rho(h) = c_{1} \frac{2^{1-\nu}}{\Gamma(\nu)} \left(\frac{h}{c_{2}}\right)^{\nu}K_{\nu}\left(\frac{h}{c_{2}}\right), \hspace{3mm} 0 \leq c_{1} \leq 1, c_{2} > 0, \nu >0,
\label{eq:wm}
\end{equation}
where $c_1$ is the nugget, $c_2$ is the range and $\nu$ is the smooth parameter.  Figure \ref{fig:schlather1} shows one realization of a process with the Whittle-Mat\'{e}rn correlation function.

To quickly summarize, we have a method for simulating realizations from max-stable processes, along with closed-form expressions for the bivariate distribution.  Max-stable processes sampled at $D$ locations form a $D-$dimensional MEVD, whose joint likelihood function can be expressed in terms of an exponent measure as in (\ref{eq:v}).  If we were to evaluate the probability that the process is below $z$ at all locations $x_1, ..., x_D$, we get
\[
P(Z(x_{1}) \leq z, ..., Z(x_{D}) \leq z) = \exp\left\{-\frac{\phi(x_{1}, ..., x_{D})}{z}\right\},
\]
where $\phi(x_{1}, ..., x_{D}) = V(1, ..., 1)$ is the \textit{extremal coefficient} for the $D$ locations.  Since the bounds on the extremal coefficient $V(z_1, ..., z_D)$ are $1/z_{1} + ... + 1/z_{D}$ and $\mbox{max}(1/z_{1}, ..., 1/z_{D})$, bounds on the extremal coefficient are $D$ and $1$, respectively, with a value of $D$ corresponding to complete independence and a value of $1$ corresponding to complete dependence.  The value can be thought of as the number of effectively independent locations among the $D$ under consideration.  For a pair of locations, the extremal coefficient is
\[
P(Z(x_{1}) \leq z, Z(x_{2}) \leq z) = \exp\left(-\frac{\phi(x_{1}, x_{2})}{z}\right).
\]
Observe that by using the bivariate distribution function available in closed form shown in (\ref{eq:schlather2}), one may write out the pairwise extremal coefficients explicitly as
\begin{eqnarray}
\phi(h) &=& 1 + \left\{\frac{1 - \rho(h; \theta)}{2}\right\}^{1/2}
\label{eq:alpha1}
\end{eqnarray}
where $h=||x_{1} - x_{2}||$.  Following \citet{smith1990max} and \citet{coles1999likelihood}, if the field $Z(\cdot)$ has been transformed to unit-Fr\'{e}chet, then $1/Z(\cdot)$ is unit exponential.  This means $1/\mbox{max}(Z(x_{1}), Z(x_{2}))$ is exponential with mean $1/\phi(x_{1},x_{2})$, and so a simple estimator of the extremal coefficient is
\begin{equation*}
\hat{\phi}(x_{1}, x_{2}) = \frac{n}{ \sum_{i=1}^n 1/\mbox{max}(z_{i}(x_{1}), z_i(x_{2}))}
\end{equation*}
where $i$ is the index for the block.  Taking this idea further, \citet{erhardt2012approximate} defined the tripletwise extremal coefficient
\begin{equation*}
P(Z(x_{j}) \leq z, Z(x_{k}) \leq z, Z(x_{l}) \leq z) = \exp\left\{-\frac{\phi(x_{j}, x_{k}, x_{l})}{z}\right\}.
\label{eq:alpha}
\end{equation*}
with estimator
\begin{equation}
\hat{\phi}(x_{j}, x_{k}, x_{l}) = \frac{n}{ \sum_{i=1}^{n} 1/\mbox{max}(z_{i}(x_{j}), z_{i}(x_{k}), z_{i}(x_{l}))}.
\label{eq:phihat}
\end{equation}

For $D$ locations the number of tripletwise extremal coefficients is ${D \choose 3}$, which grows quite rapidly as $D$ increases.  As part of an ABC analysis, lower dimensional summary statistics are preferred.  Hence, a clustering step can be added to group these coefficients into a fixed number $K << {D \choose 3}$ (see \citet{erhardt2012approximate} for full details on this step).
The ${D \choose 3}$ triplet extremal coefficients may be estimated for the observed data using (\ref{eq:phihat}), and then these values are averaged within the $K$ clusters.  The result is the summary of the observed data, $s = (\bar{\phi}_{1}, ..., \bar{\phi}_{K})$.  

Following Algorithm \ref{alg:ABCimportanceSampling}, independent draws from the prior $\theta^{(i)} \sim p(\theta)$ are taken.  For each draw from the prior, a max-stable process $Z^{(i)}$ with unit-Fr\'{e}chet margins is simulated on the same locations and for the same number of years as the observed data.  For this data $Z^{(i)}$, all tripletwise extremal coefficients are estimated, and averages within each cluster group are taken to produce $s^{(i)} = (\bar{\phi}_{1}, ..., \bar{\phi}_{K})$.  To compare summaries, the sum of absolute deviations is used a distance measure, so that
\begin{equation*}
	\|s^{(i)}-s_{obs}\| = \sum_{k=1}^{K} |s^{(i)}_{k} - s_{k,obs}|
\label{eq:d}
\end{equation*}
where $s^{(i)}_{k}$ and  $s_{k,obs}$ denote the $k$-th element of $s^{(i)}$ and $s_{obs}$ respectively. 
As before, the smoothing kernel $K_h(u)$ is uniform over $(-h, h)$, and the kernel scale parameter $h$ is determined as the 0.01 quantile of the $N$ samples $\|s^{(i)}-s_{obs}\|$.

%%%%%%%%%%%%%
\subsection{Application}%
%%%%%%%%%%%%%

The observed data are annual maxima daily temperatures taken from 39 locations in the midwestern United States with complete summer (June 1 - August 31) temperature records from 1895 to 2009.  All sites are located between longitudes 93 and 103 degrees west, and latitudes 37 to 45 degrees north.

Each of the 39 marginal distributions were transformed to unit-Fr\'{e}chet through ordinary maximum likelihood estimation.  Based on the dependence model (\ref{eq:wm}) with no nugget effect ($c_1=1$), the prior for the spatial dependence parameter  vector $\theta = (c_2, \nu)$ is specified as $[0,7] \times [0,7]$.  This choice aims to represent vague non-informativeness and also place positive support on a large range of possible parameter values.  Based on candidate draws $\theta^{(i)}\sim p(\theta)$, $i=1,\ldots,N$, max-stable processes $Z^{(i)} \sim f(Z | \theta^{(i)})$ were simulated for 115 years at the same 39 locations.

\begin{figure}
\begin{center}
\includegraphics[width=3.5in]{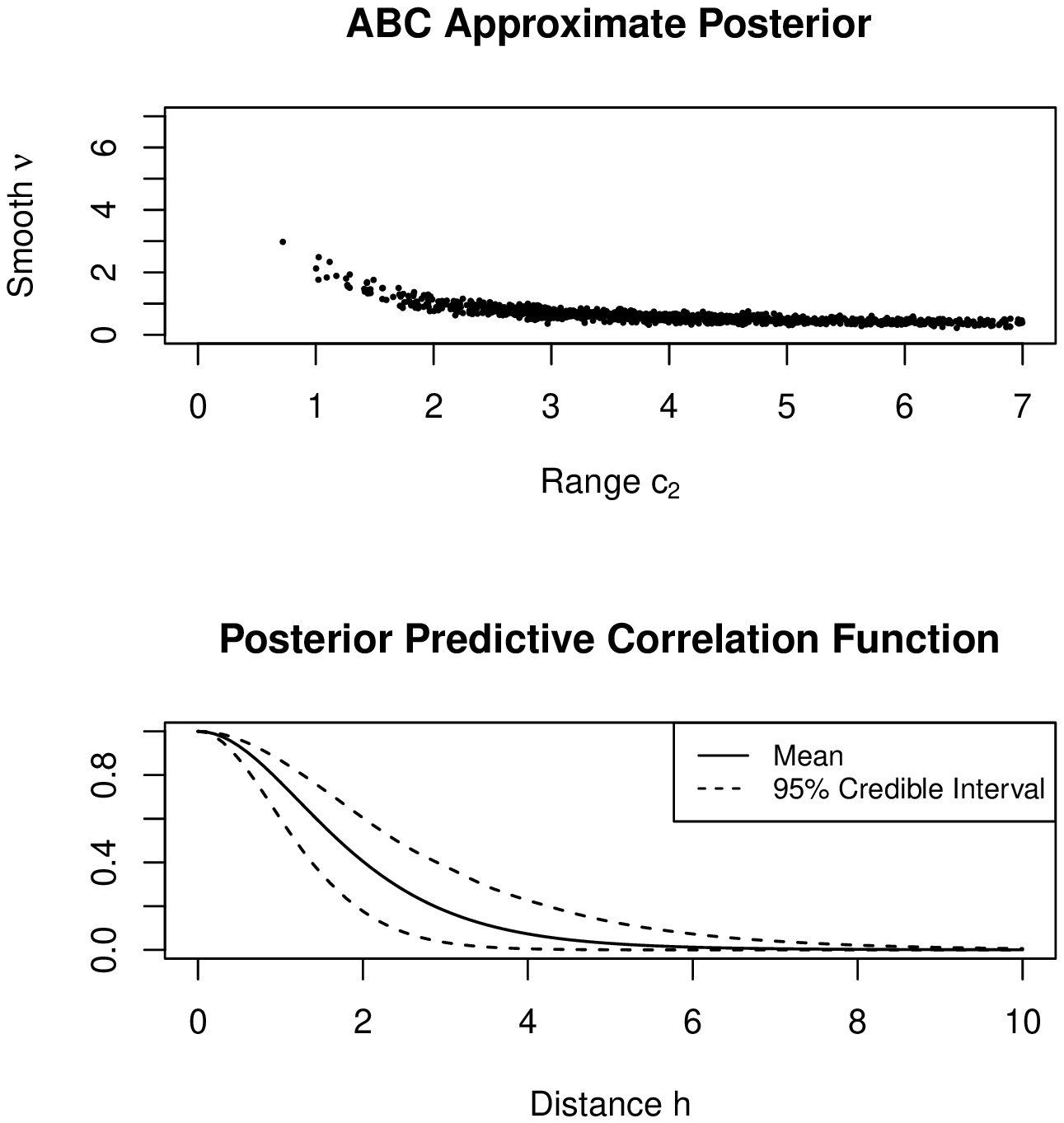}
\caption[ABC posterior of the dependence parameters]{Top: Samples drawn from the ABC approximate posterior distribution based on fitting a max-stable process.  The uniform prior for $(c_2,\nu)$ is the full range displayed on the scatterplot.  Bottom: Posterior predictive distribution of the corresponding spatial correlation function $\rho(h)$ following equation~(\ref{eq:wm}).  The solid line shows the pointwise posterior mean, and the dashed lines show the pointwise 2.5\% and 97.5\% quantiles.}
\label{fig:abcpost}
\end{center}
\end{figure}

The ABC posterior approximation $\pi_h^{ABC}(\theta|y_{obs})$ based on $N=100,000$ samples is shown in Figure~\ref{fig:abcpost}.  The approximate posterior has correctly identified the credible regions of the parameter space from these data, retaining only draws from a crescent shaped subspace.  Inversely changing values of $c_2$ and $\nu$ can produce similarly shaped correlation functions $\rho(h)$ -- hence the crescent shape for the posterior approximation.  The posterior predictive distribution of spatial correlation functions $\rho(h)$  is illustrated in the lower panel of Figure \ref{fig:abcpost}.  This approximate posterior may now be used to fully incorporate the parameter uncertainty of $\theta$ into estimates of the distribution of total loss $P = P_1 + \ldots  +P_D$ and associated actuarial risk measures.

%%%%%%%%%%%
%%%%%%%%%%%%
\section{Discussion}%%
%%%%%%%%%%%%
%%%%%%%%%%%

This chapter has outlined how approximate Bayesian statistical modeling of extremes can be viable in situations where the likelihood function either cannot be evaluated numerically, or written in closed form.   Sections \ref{sec:stereo} and \ref{sec:maxstab}  illustrate ABC methods as applied to  two different models in extremes, with various model and data complexities, in which an intractable likelihood is a common issue.
All that ABC methods require in order to be implemented is a fast algorithm to simulate realisations from the intractable likelihood, and a suitable vector of summary statistics.
The former requirement is not always a trivial procedure for extremal models, and commonly needs some subtle theory and computational tricks. However, procedures for accurately simulating observations from extremal models continues to be an area of active research.

Perhaps the most important aspect of any ABC analysis is the identification of a low-dimensional vector of summary statistics that are highly informative for the model parameters. Once the summary statistic is chosen, the best possible ABC approximation to the posterior is $\pi(\theta|s_{obs})\approx\pi(\theta|y_{obs})$, and this occurs when the kernel scale parameter $h\rightarrow 0$. If the summary statistic is poorly chosen, the rest of the ABC procedure (e.g. Algorithm \ref{alg:ABCimportanceSampling}) can not recover the loss of information. In fact, as $h>0$ is a necessity in practice, the best ABC posterior approximation achievable is $\pi_h^{ABC}(\theta|s_{obs})$ which is in general a worse approximation than $\pi(\theta|s_{obs})$. Hence, the choice of summary statistics is critical. To date, probably the most useful and principled general approach to identify summary statistics is developed by \citet{fearnhead+p12}, whereas a method to derive summary statistics from the score function of a related composite model (such as a pairwise likelihood) could be of particular interest for modelling spatial extremes \citep{ruli+sv13}.

Coupled with the choice of summary statistics is the issue of computational overheads. ABC methods are a classic example of a computation versus accuracy trade-off. This implies that any developments in improving the efficiency of an ABC analysis can be converted into producing a more accurate inference for the same computational cost. Some more sophisticated algorithms than importance sampling are detailed in Section \ref{sec:otherABC}, but other ideas, such as terminating the generation of a dataset early if it is likely to be rejected \citep{prangle14} are also starting to be developed. Improving the ``approximation'' (in Approximate Bayesian Computation) while reducing the ``computation'' is an active research area in ABC.

It is worth stating that the development of a tractable, closed-form joint likelihood for a stochastic process would entirely obviate the need for ABC methods by allowing for a more conventional Bayesian analysis based on direct usage of the likelihood.  A flavour of this evolution can be found in both \citet{genton2011likelihood} and \cite{aandahl+sst14}.  After a period of roughly two decades in which researchers could write down only the bivariate joint distribution function for the Gaussian max-stable process, \citet{genton2011likelihood} extended the closed-form to include the trivariate joint likelihood. In the setting of population genetics,  \cite{aandahl+sst14} show that models that were previously only available for analysis using ABC, have subsequently become tractable with further analytical development. These are  useful reminders that what is considered intractable today may not remain intractable in the future. In the meantime, however, the statistician's toolbox remains empowered through the availability of ABC methods.

\bibliographystyle{chicago}
\bibliography{abc-extremes}

\end{document}